\documentclass[11pt,twoside]{article} 
\usepackage{asp2004}
\usepackage{epsf}
\usepackage{psfig}
\usepackage{lscape}
\usepackage{graphicx}
\usepackage{natbib}
\usepackage{multirow}
\usepackage{verbatim}
\begin{document}
\title{Discovery of a Long-Period Photometric Variation in the
  V361~Hya Star HS\,0702+6043}
\author{S.~Schuh$^1$,  
  J.~Huber$^1$, 
  E.M.~Green$^2$,
  S.J.~O'Toole$^3$,
  S.~Dreizler$^1$,
  U.~Heber$^3$,
  G.~Fontaine$^4$
}
\affil{$^1$Institut f\"ur Astrophysik, Universit\"at G\"ottingen,
  Friedrich-Hund-Platz\,1, D--37077~G\"ottingen, Germany\\
  $^2$Steward Observatory, University of Arizona,
  933~N~Cherry~Ave., Tucson~AZ~85721-0065, USA\\
  $^3$Dr. Remeis-Sternwarte Bamberg, Universit\"at
  Erlangen-N\"urnberg, Sternwartstra\ss e 7, D--96049
  Bamberg, Germany\\
  $^4$D{\'e}partement de Physique, Universit{\'e} de Montr{\'e}al, C.P.~6128,
  Succursale Centre-Ville, Montr{\'e}al, QC~H3C~3J7, Canada}
\begin{abstract} 
  We report the discovery of a long-period g-mode oscillation in the
  previously known short-period p-mode sdB pulsator HS\,0702$+$6043. This
  makes this star an extraordinary object, unique as a member of the
  family of sdB pulsators, and one of the very few known pulsating stars
  overall exhibiting excited modes along both the acoustic and gravity
  branches of the nonradial pulsation spectrum. Because p-modes and
  g-modes probe different regions of a pulsating star, HS\,0702$+$6043 holds
  a tremendous potential for future detailed asteroseismological
  investigations.
\end{abstract}
\section{SdB Asteroseismology: V361~Hya and lpsdBV Stars}
Subluminous B (sdB) stars dominate the population of faint blue stars
of our own galaxy down to m$_V\approx$\,16 and are found in the disk
(field sdBs) as well as in globular clusters. 
SdB stars can be identified with models of the Extended Horizontal
Branch (EHB) burning He in their core \citep{heber:86}.
Photometric variations due to radial and non-radial pulsations in some
of the sdB stars (V361~Hya stars, formerly referred to as EC\,14026 or
sdBV stars, 33 are known today) were detected by \citet{kilkenny:97}.
The periods of these stars are found to be in the range 80\,--\,600\,s
with low amplitudes (few mmag). Log\,${g}$ and $T_{\rm eff}$ for the
pulsators range from 5.3\,--\,6.1\,dex and
29\,000\,--\,36\,000\,K. The pulsations are driven by an opacity bump
due to mainly Fe in subphotospheric layers
\citep{1997ApJ...483L.123C}. First successes in asteroseismological
analyses of sdBVs were reported by \citet{2001ApJ...563.1013B} and
Charpinet et al.\ (these proceedings).
A new class of multi-mode sdB pulsators has been discovered by
\citet[][25 so-called \mbox{lpsdBV} known today, for which
  PG\,1716$+$426 is the prototype]{2003ApJ...583L..31G}. Their pulsation
periods are about 10 times longer ($\approx$\,1\,h) than those of the
V361~Hya stars.
\begin{table}[!t]
  \caption[]{Data log of photometric observations obtained for
    HS\,0702+6043.}
  \smallskip
  \begin{center}
    \label{tab:schuh1}
	  {\small
	    \begin{tabular}{lrlr@{/}c@{/}lr@{:}lr@{:}lr}
	      \noalign{\smallskip}
	      \tableline
	      \noalign{\smallskip}
	      Observatory&Telescope&Filter&\multicolumn{3}{l}{UT date}&
	      \multicolumn{2}{c}{start}&\multicolumn{2}{c}{stop}&cycle time\\
	      &&&[y&m&d]&[h&m]&[h&m]&[s]\\
	      \noalign{\smallskip}
	      \tableline
	      \noalign{\smallskip}
	      Calar Alto&1.2\,m&\textit{none}&1999&12&07&01&19&06&16&18\\
	      Calar Alto&1.2\,m&\textit{none}&1999&12&08&04&28&06&23&16\\
	      Calar Alto&1.2\,m&\textit{none}&1999&12&09&04&46&06&09&21\\
	      T\"ubingen&0.8\,m&G(green)&2004&02&04&19&55&03&12$^*$&38\\
	      Steward&2.3\,m&F555W&2004&02&09&02&36&08&25&86\\
	      Steward&2.3\,m&F555W&2004&02&10&02&40&08&55&86\\
	      \noalign{\smallskip}
	      \tableline
	      \noalign{\smallskip}
	      \multicolumn{11}{r}{\hfill$^*$implies date change between
		start and stop}
	    \end{tabular}
	  }
  \end{center}
\end{table}
Hence the sdB variables come in two flavours, the V361~Hya and the
\mbox{lpsdBV} stars. The longer periods (lp) in the \mbox{lpsdBV}
imply the presence of excited high radial order gravity modes
(g-modes), in contrast to the pressure modes (p-modes) present in the
V361~Hya stars. The driving mechanism for these new pulsators is under
discussion. \citet{2003ApJ...597..518F} favour the iron opacity bump
meachnism, but also consider excitation by tidal forces in close
binaries a viable alternative.  Constructing pulsation models based on
such physical insights is a necessary prerequisite to interpret
asteroseismological observations. The internal structure obtained from
such modeling allows to extract clues on the evolutionary history of
sdBs which is still much under debate, and constrains the future
evolution: sdB stars evolve directly into white dwarfs and hence
represent one of the feeder channel to white dwarfs.
\section{Frequency analysis of HS\,0702+6043}
\begin{figure}[!t]
  \plotone{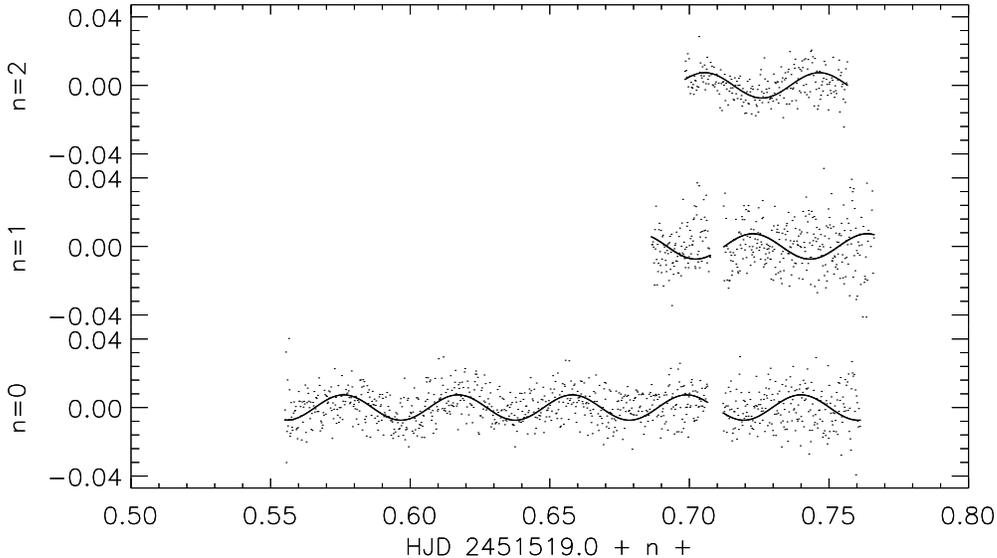}
  \vspace{-15mm}
  \caption{Calar Alto 1.2\,m light curves (individual data points) from
  December 1999 with two short-period oszillations ($f_1$ and
  $f_2$) subtracted; the y-axis unit is in mi. Overlaid (full line) is
  a sine curve corresponding to $f_3$.}
  \label{fig:schuh2}
\end{figure}

\begin{figure}[!t]
  \plottwo{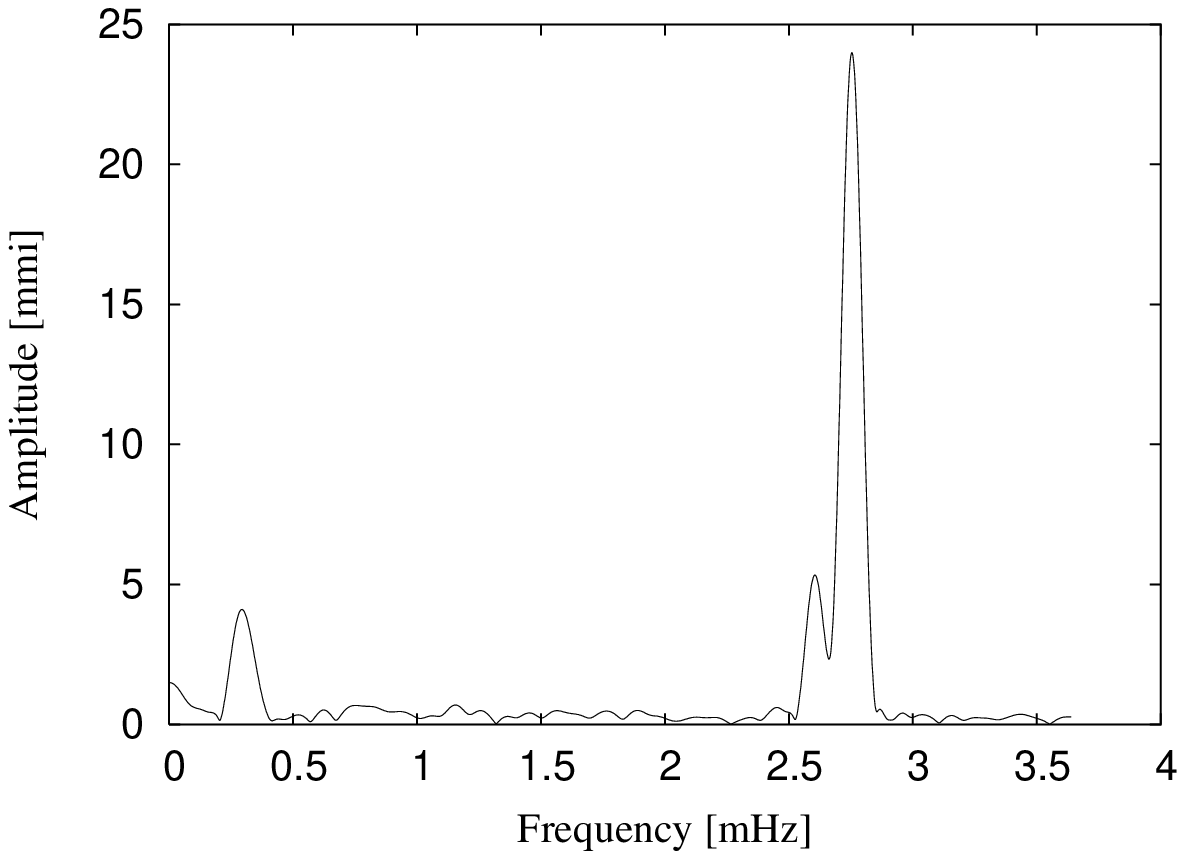}{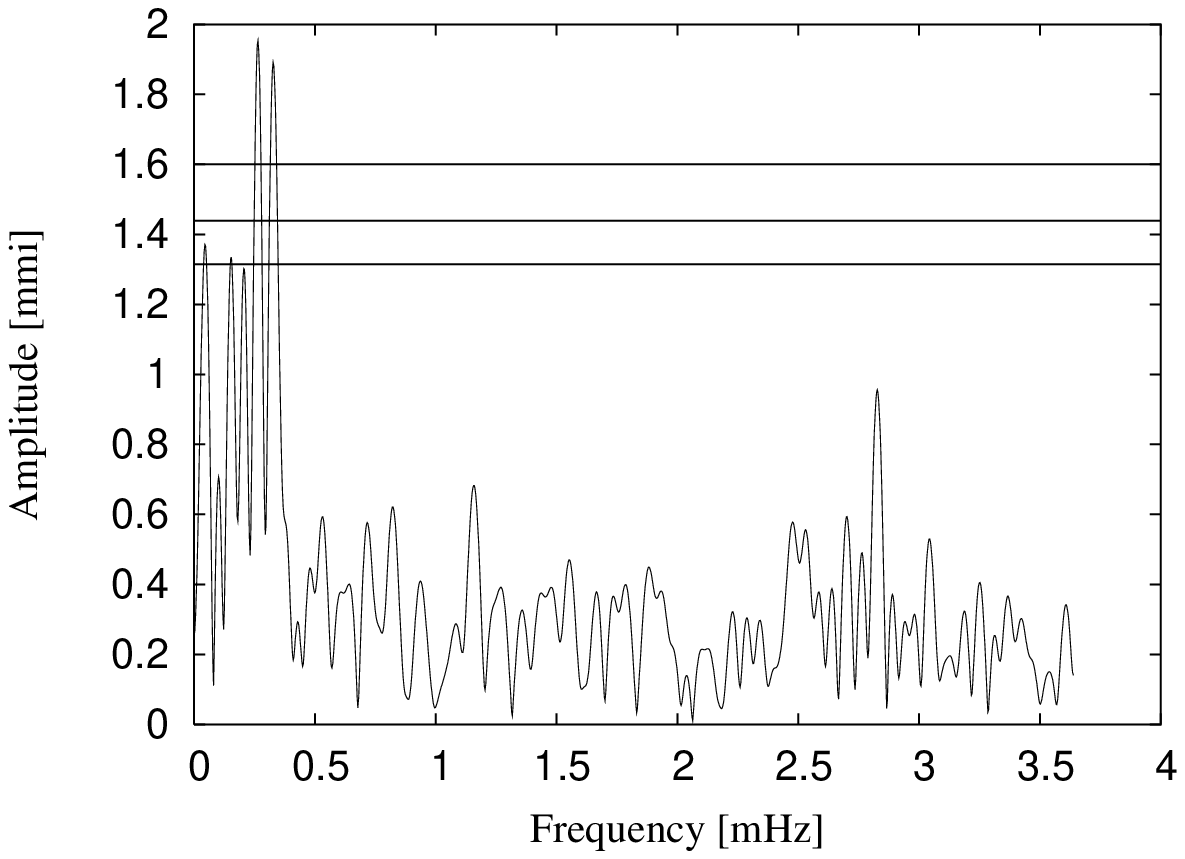}
  \caption{Left panel: Fourier spectrum of one 6\,h-observation of
    HS\,0702$+$6043 with the 2.3\,m Steward telescope at Kitt Peak: the 
    three main modes from the full light curve are clearly
    visible. Right panel: Fourier spectrum of the same light curve
    after substraction of the three modes as listed 
    in Table~\ref{tab:schuh2}: There are remaining amplitudes well
    above the significance levels expected of spectra
    from time series conforming to the hypothesis 
    of independent identically distributed noise (from bottom to top,
    the horizontal lines mark false alarm probabilities of 0.05, 0.01
    and 0.001).\label{fig:schuh3}}
\end{figure}

In follow-up observations of hot subdwarfs from the Hamburg Quasar
survey, the m$_B=$\,15 object HS\,0702+6043 was discovered as a new
V361~Hya star by \citet{2002A&A...386..249D}. The main pulsation
period was 363\,s at a relatively large amplitude of 29\,mmag. A second
period of 382\,s was present at a much smaller amplitude of 3.8\,mmag.
The atmospheric parameters derived by quantitative model atmosphere
analysis, $T_{\rm eff}=$\,28\,400\,K and $\log{g}=$\,5.35, make it one
of the coolest V361~Hya stars known.
With refined time series analysis tools, we recently discovered in the
original data set that HS\,0702$+$6043 shows long period variations of
about 1\,h (at a much lower amplitude) along with the short period
oscillations. The initial discovery was confirmed by observations
with the 2.3\,m Steward telescope atop Kitt Peak in February 2004.
\begin{figure}[!t]
  \plotone{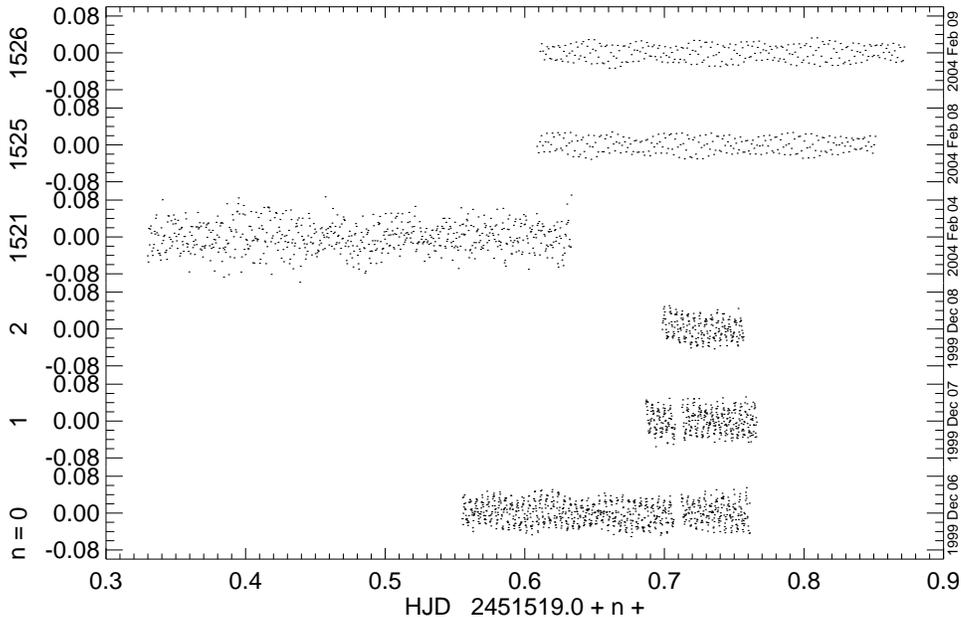}
  \vspace{-10mm}
  \caption{Photometric observations obtained for
    HS\,0702$+$6043; y-axis is in mi
    (modulation intensity, $\Delta I/\bar{I}$)
.}
  \label{fig:schuh1}
\end{figure}
Table~\ref{tab:schuh1} lists the photometric observations available
for HS\,0702$+$6043 so far. Figure~\,\ref{fig:schuh1} shows the
corresponding light curves. An analysis of the short periods in the
1999 Calar Alto data set has been published in
\citet{2002A&A...386..249D}. In a re-analysis of these data, a third
period of about 1\,h started to emerge (Fig.\,\ref{fig:schuh2}), which
prompted us to initiate follow-up observations.
The data obtained in T\"ubingen allow to recover the strongest mode at
363\,s only. As the telescope is located inside the city limits of
T\"ubingen, we regard this detection as a success. Of course, the nearly
noise-free Steward data are much better suited for our aim. The
two consecutive nights not only allow to extract the two short periods
previously published but also one mean long period
(Table~\ref{tab:schuh2}). In addition they 
indicate a more complicated structure of the low-frequency variation
(Fig.\,\ref{fig:schuh3}). Thus, the existence of the long period is
clearly confirmed; it is still present four years after the first
observation. It can be detected in both nights of the new data set
already in the frequency spectrum of the full light curve without
prewhitening. We show the spectrum of the first night in
Fig.\,\ref{fig:schuh2}. After prewhitening of the three periods listed
in Table~\ref{tab:schuh2}, there are still significant residual
amplitudes, implying the presence of more than one period in the
low-frequency range (again, this is true for both nights).
This finding strongly argues for a pulsational character of the new
low-amplitude variation. It also supports the view that the new
frequency is not simply a higher-order combination frequency as might
be inferred from Table~\ref{tab:schuh2} due to the consistency of the
difference within the errors, but rather corresponds to an independent
mode. The complete absence of all further combination frequencies
that could be expected at a detectable level also dismisses this
suggestion.
\begin{table}[!hb]
  \caption[]{Frequencies found from February 2004 Steward 2.3\,m 
    telescope data.}
  \smallskip
  \begin{center}
    \label{tab:schuh2}
	  {\small
	    \begin{tabular}{lrrrr}
	      \noalign{\smallskip}
	      \tableline
	      \noalign{\smallskip}
	      &frequency &period &amplitude&$\Delta$f\\
	      &[$\mu$Hz]&[s]&[mmi]&[$\mu$Hz]\\
	      \noalign{\smallskip}
	      \tableline
	      \noalign{\smallskip}
	      $f_1$&2754\,(9) &  363&   21.7(5) &\\
	      $f_2$&2617\,(9) &  382&    4.6(5) &\\
	      $f_3$& 283\,(9) & 3538&    3.7(5) &\quad\quad 2($f_1-f_2)=$274\\
	      \noalign{\smallskip}
	      \tableline
	    \end{tabular}
	  }
  \end{center}
\end{table}
\clearpage
\section{Interpretation of the New 1\,h Period as a g-Mode Pulsation}
\begin{figure}[!t]
  \plottwo{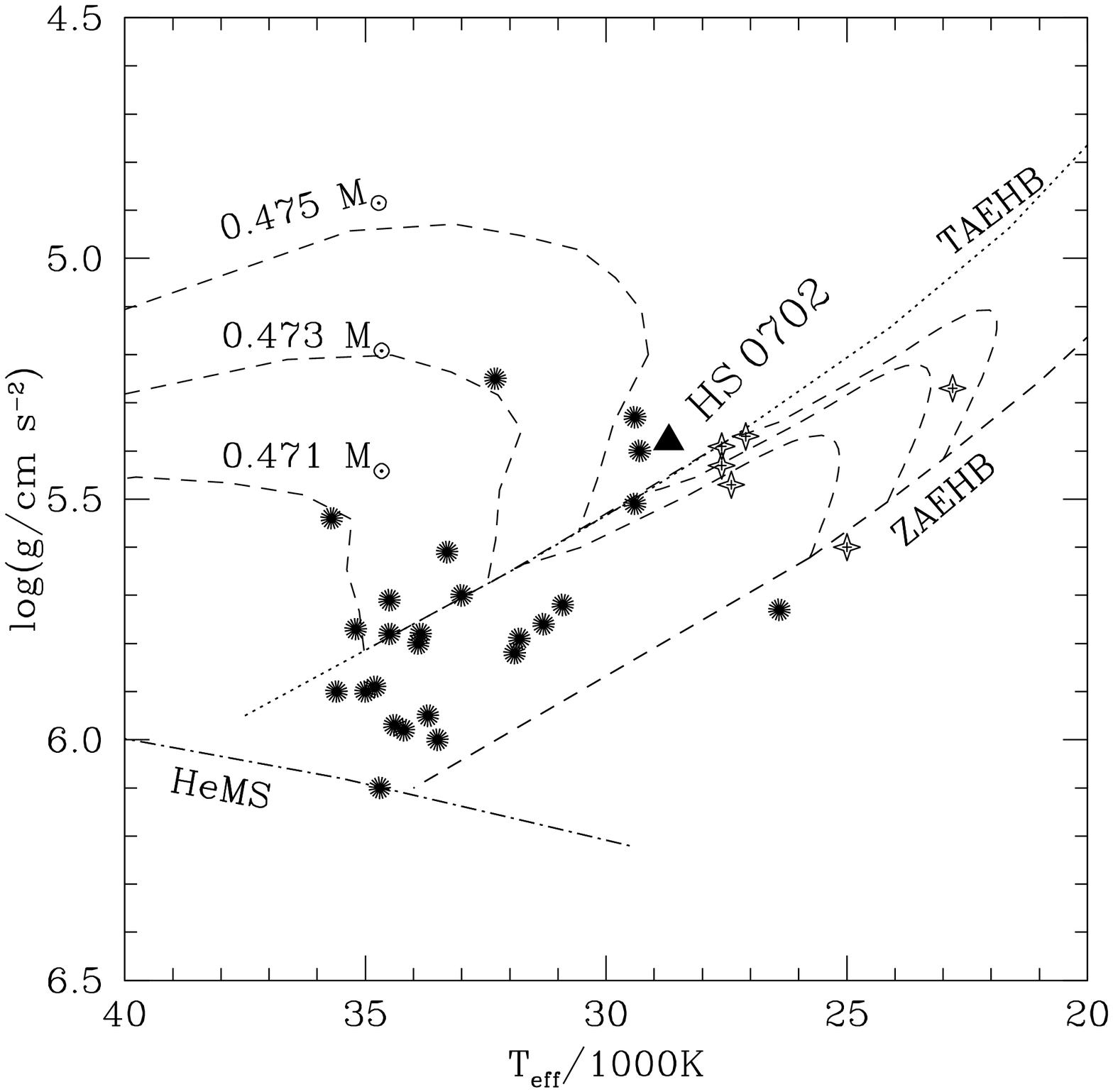}{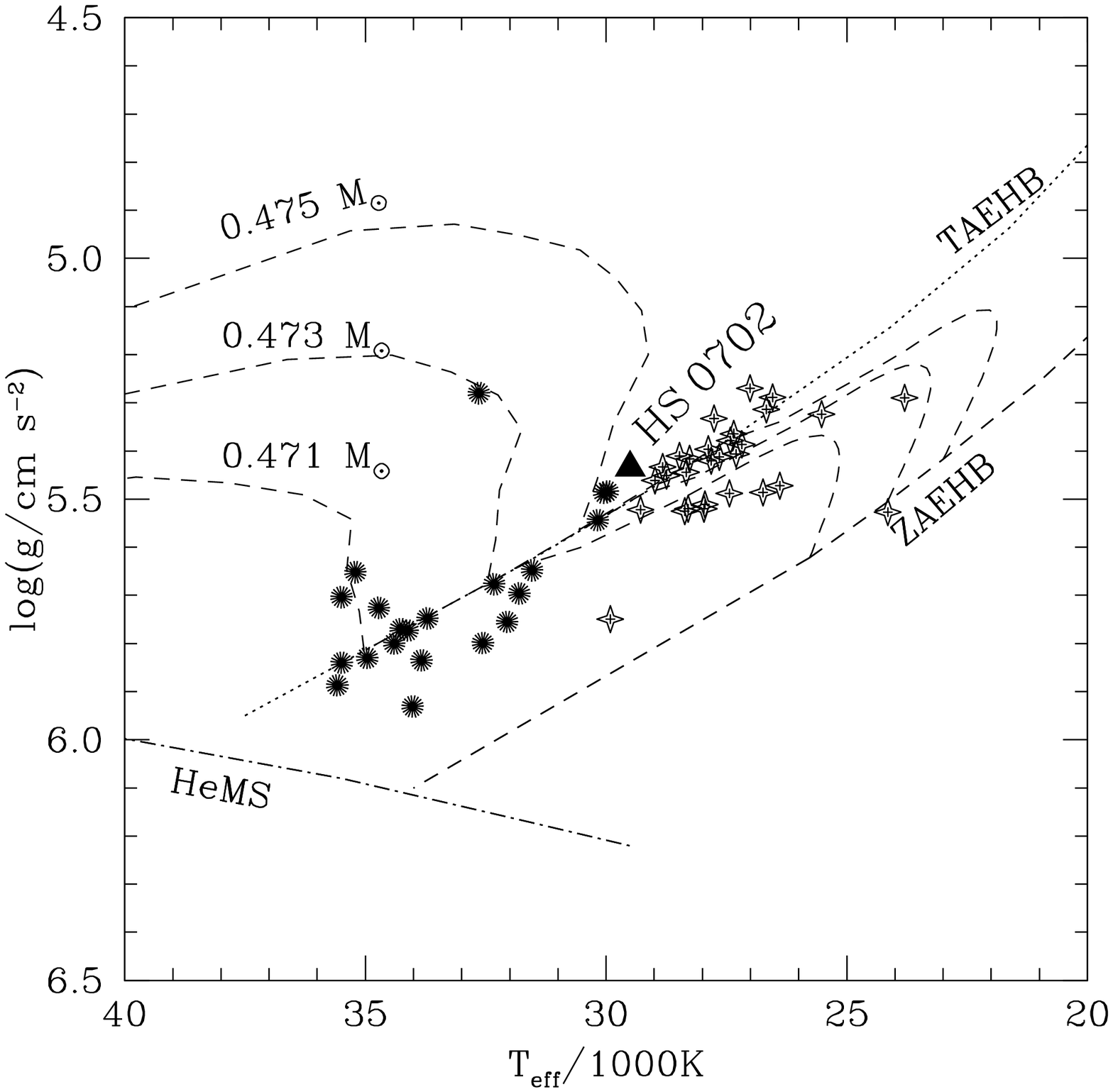}
  \caption{Known sdBV and lpsdBV pulsators (all published parameters:
    left, all observed by E.M.G.: right) in the $\log{g}-{\rm T_{\rm eff}}$
    diagram: sdBV locations are identified through asterisks, lpsdBV
    locations by square stars. The helium main sequence,
    zero and terminal age extended horizontal branches and HS0702+6043 are
    marked; also shown are evolutionary tracks off the extended
    horizontal branch by \citet{dorman:95}.\label{fig:schuh4}}
\end{figure}

Both the relative shortness of an hypothetical orbital or rotational period
as well as the multi-mode character of the photometric variation make
a successful explanation through binary or rotational motion highly
unlikely. On the other hand, it is known that g-mode pulsators with
similar periods and amplitudes exist among sdBs with parameters close
to those found for HS\,0702$+$6043: its effective temperature places
it at the cool end of the p-mode, and hence at the same time close to
the hot end of the g-mode, empirically determined instability regions.
This strongly suggests that p-mode and high-order g-mode pulsations
coexist in this star, indicating that the instability strip for
\mbox{lpsdBV} stars might overlap with that for V361~Hya stars.
Whether or not the instability strips for \mbox{lpsdBV} and V361~Hya
stars overlap is an important question. In Fig.\,\ref{fig:schuh4}
we plot the published atmospheric parameters of 27 V361~Hya stars and 6
lpsdBVs \citep{2001MNRAS.326.1391M} indicating that the two instability regions
do indeed touch. However, a homogenous set of accurate atmospheric
parameters is still needed. In a first attempt, which in particular includes more
lpsdBV stars, spectroscopic work by two of us (E.M.G.\ and G.F., see
Fig.\,\ref{fig:schuh4}) has homogenized the parameter determination to
obtain comparable temperatures and gravities. This
analysis yields spectroscopic parameters of $T_{\rm eff}=$\,29\,500\,K
and $\log{g}=$\,5.44 for HS\,0702$+$6043 that differ from the earlier
results. Despite systematic shifts introduced by neglecting metal-line
blanketing effects, the relative position of HS\,0702$+$6043 confirmes the
conclusion that the two instability regions touch. This position makes
HS\,0702$+$6043 a key object in (not only sdB) asteroseismology since
simultaneous p- and g-mode pulsations sample different parts of a
star.
\citet{2003ApJ...597..518F} suggested that the sdBV/lpsdBV have
a main sequence analogy in the ${\beta}$~Cep/[SPB] variables. In
this context, it is interesting to note that, according to
\citet{2004MNRAS.347..454H}, the $\beta$~Cep star $\nu$~Eri might also
show an [SPB]-like frequency. Similarly, \citet{2002MNRAS.333..262H}
have reported a possibly $\delta$~Sct-like frequency in the
$\gamma$~Dor object HD~209295.
With HS\,0702$+$6043 belonging to both sdBV classes, it challenges
theory's description of stability and driving mechanisms in current
pulsational models. We have constructed a preliminary stellar
structure pulsational model for HS\,0702$+$6042 where both
short-period, low-order p-modes and long-period, high-order g-modes
can be excited. This is in good qualitative agreement with what we
observe in HS\,0702$+$6043, but the effective temperature of that
model is somewhat lower than desirable. Alternatively, we propose that
those g-modes whose frequencies happen to fall close to nonlinear
peaks from p-modes (combination frequencies) could be excited through
resonance coupling.  When the star is ''hit'' at a frequency $f_2-f_1$
at relatively large amplitudes, g-modes which happen to have very
similar frequencies (if they exist), and which are normally stable,
would then be driven to observable amplitude because they resonate
with the ''engine'' that hits them. This would also imply that the
low-frequency peaks are real g-modes excited through resonance at
those frequencies and not merely nonlinear features.
\section{Summary and Outlook}
We report the recent discovery of a new period in the sdB
star HS\,0702$+$6043 already known as a V361~Hya variable. The
additional detection is a low-amplitude, long-period light variation
that is most probably due to independent pulsations. The
interpretation as a member of the recently discovered class of g-mode
pulsators among sdB stars makes this variable the first recognized to
show both of the known types of pulsational variations for sdBs.
The current light curves prove that beyond the one g-mode type period
confirmed so far, there is residual power in the low-frequency regime,
suggesting a more complicated structure. We plan to obtain a good
frequency spectrum from a longer light curve to further investigate
the g-modes, and adequately resolve the low-frequency
spectral structure.
There are clues that HS\,0702$+$6043 might in fact represent the
prototype of a whole new class of similar objects: rumours at this
conference suggest that the interesting behaviour observed in
HS\,0702$+$6043 may also have been uncovered in other V361~Hya
stars, but this remains to be confirmed.
\acknowledgements{We thank S.\,Charpinet, R.\,Silvotti
  and A.\,Baran for interesting discussions.
  S.\,Schuh thanks the conference organizers for financial support.
  The 1999 data were collected at the
  German-Spanish Astronomical Center (DSAZ), Calar Alto, operated by
  the Max-Planck-Institut f\"ur Astronomie Heidelberg jointly with
  the Spanish National Commission for Astronomy. 
  T.\,Nagel contributed data using the T\"ubingen 80\,cm telescope.
} 
\end{document}